\documentclass[12pt, a4paper]{article}
\usepackage{graphicx}
\usepackage{amssymb}
\usepackage{amsmath}
\usepackage{bm}
\usepackage{color}
\usepackage{theorem}
\usepackage{subcaption}
\usepackage{listings}
\usepackage{colortbl}
\usepackage{tabularx}
\usepackage{longtable}
\usepackage[utf8]{inputenc}
\usepackage[T1]{fontenc}
\usepackage{lmodern}
\usepackage[top=30truemm,bottom=30truemm,left=25truemm,right=25truemm]{geometry}

\usepackage[sort&compress,numbers, merge]{natbib}

\definecolor{Orange}{cmyk}{0,0.61,0.87,0}
\definecolor{JungleGreen}{cmyk}{0.99,0,0.52,0}
\definecolor{OliveGreen}{cmyk}{0.64,0,0.95,0.40}
\definecolor{Brown}{cmyk}{0,0.81,1,0.60}
\definecolor{RoyalBlue}{cmyk}{0.71,0.53,0,0.12}
\definecolor{Gray}{cmyk}{0,0,0,0.40}
\definecolor{LightPink}{cmyk}{0.0,0.25,0,0}
\definecolor{LLightPink}{cmyk}{0.0,0.10,0,0}
\definecolor{LightBlue}{cmyk}{0.25,0,0,0}
\definecolor{LightGray}{cmyk}{0,0,0,0.2}

\allowdisplaybreaks[1]

\usepackage[colorlinks=true, linkcolor=OliveGreen, citecolor=RoyalBlue,
urlcolor=RoyalBlue]{hyperref}

\renewcommand{\thefootnote}{\fnsymbol{footnote}}

\begin{document}

\begin{titlepage}

  \begin{flushright}

\end{flushright}

\vskip 1.5cm
\begin{center}

{\large
{\bf
Axion Emission from Proton Cooper Pairs in Neutron Stars 
}
}

\vskip 1.5cm

Koichi~Hamaguchi$^{a,b}$\footnote{
  E-mail address: \href{mailto:hama@hep-th.phys.s.u-tokyo.ac.jp}{\tt hama@hep-th.phys.s.u-tokyo.ac.jp}}, 
Natsumi Nagata$^a$\footnote{
E-mail address: \href{mailto:natsumi@hep-th.phys.s.u-tokyo.ac.jp}{\tt natsumi@hep-th.phys.s.u-tokyo.ac.jp}},
and
Jiaming~Zheng%
\footnote{
  E-mail address: \href{mailto:zhengjm3@gmail.com}{\tt zhengjm3@gmail.com}}

\vskip 0.8cm

{\it $^a$Department of Physics, University of Tokyo, Bunkyo-ku, Tokyo
 113--0033, Japan} \\[2pt]
 {\it $^b$Kavli IPMU (WPI), University of Tokyo, Kashiwa, Chiba
  277--8583, Japan} \\[2pt]

\date{\today}

\vskip 1.5cm

\begin{abstract}

  We investigate axion emission from singlet proton Cooper pairs in neutron stars, a process that dominates axion emission in young neutron stars in the KSVZ model. By re-deriving its emissivity, we confirm consistency with most existing literature, except for a recent study that exhibits a different dependence on the effective mass. This discrepancy results in more than an order-of-magnitude deviation in emissivity, significantly impacting constraints on the KSVZ axion from the cooling observations of the Cassiopeia A neutron star. Furthermore, we examine uncertainties arising from neutron-star equations of state and their role in the discrepancy, finding that the large deviation persists regardless of the choice of equations of state.

\end{abstract}

\end{center}
\end{titlepage}

\renewcommand{\thefootnote}{\arabic{footnote}}
\setcounter{footnote}{0}

\section{Introduction}

Axion~\cite{Weinberg:1977ma, Wilczek:1977pj}\footnote{For a recent review on axions, see Refs.\cite{DiLuzio:2020wdo, Choi:2020rgn, Antel:2023hkf, ParticleDataGroup:2024cfk}.} is a pseudo-Nambu-Goldstone boson arising from the spontaneous breaking of the Peccei-Quinn symmetry~\cite{Peccei:1977hh, Peccei:1977ur}, originally proposed to resolve the strong CP problem. Its coupling to gluons via the quantum anomaly of the Peccei-Quinn symmetry induces an effective potential for the axion field, driving the system toward a CP-conserving vacuum. Several concrete realizations of this mechanism have been proposed in the literature, among which the KSVZ~\cite{Kim:1979if, Shifman:1979if} and DFSZ~\cite{Zhitnitsky:1980tq, Dine:1981rt} models are particularly well-known. In these models, the Peccei-Quinn symmetry-breaking scale is significantly higher than the electroweak scale. Since the couplings of axions are inversely proportional to this scale—a common feature of Nambu-Goldstone bosons—their interactions with Standard Model particles are extremely weak, and their mass, generated by the effective potential, is very small. These characteristics make experimental searches for axions highly challenging.

Observations of astrophysical objects place the most stringent constraints on axion models~\cite{Caputo:2024oqc, Carenza:2024ehj}. In particular, SN1987A has provided the strongest limits over the years~\cite{Raffelt:2006cw}. Axions, being extremely light and weakly coupled to ordinary matter, can freely escape from a core-collapse supernova once produced, whereas neutrinos remain trapped immediately after the collapse. Consequently, if too many axions are emitted, the supernova cools too rapidly, shortening the duration of neutrino emission compared to that observed for SN1987A. This observation thus restricts the axion-nucleon couplings. For example, a recent analysis constrains the axion decay constant to $f_a \gtrsim 4 \times 10^8$~GeV in the KSVZ model~\cite{Carenza:2019pxu} (see also Refs.~\cite{MartinCamalich:2020dfe, Carenza:2020cis, Camalich:2020wac, Choi:2021ign, Vonk:2022tho, Ho:2022oaw, Lella:2023bfb, Cavan-Piton:2024ayu}).

A cooling constraint on axion-nucleon couplings can also be obtained from neutron star (NS) temperature observations~\cite{Iwamoto:1984ir, Umeda:1997da, Leinson:2014ioa, Sedrakian:2015krq, Paul:2018msp, Hamaguchi:2018oqw, Beznogov:2018fda, Leinson:2021ety, Buschmann:2021juv, Gomez-Banon:2024oux}. Currently, the temperatures of approximately 60 NSs have been measured~\cite{Potekhin:2020ttj, tempdata}, and the observed data generally align with the predictions of standard NS cooling theory~\cite{Yakovlev:1999sk, Yakovlev:2000jp, Yakovlev:2004iq, Page:2004fy, Gusakov:2004se, Page:2009fu}.\footnote{However, some older NSs exhibit significantly higher temperatures than those predicted by the standard cooling scenario~\cite{Kargaltsev:2003eb, Mignani:2008jr, Durant:2011je, Rangelov:2016syg, Pavlov:2017eeu, Abramkin:2021tha, Abramkin:2021fzy}. These observations can be explained by internal heating mechanisms~\cite{Gonzalez:2010ta, Yanagi:2019vrr, Fujiwara:2023tmr}, which are relevant only for old NSs and, in particular, do not affect the temperature of young NSs.} Given this, NS temperature data can be used to constrain the existence of additional cooling sources, such as axions. Young NSs are particularly useful for this purpose, as the axion emission rate rapidly increases with temperature.

As it turns out, a young NS in the supernova remnant Cassiopeia A (Cas A) is potentially one of the most sensitive probes of axions. Since 2006, Chandra observations have revealed a steady decline in the surface temperature of the Cas A NS~\cite{Heinke:2010cr}.  For a long time, this cooling trend was primarily observed in data taken in the GRADED mode of the ACIS detectors, which may suffer from pile-up effects~\cite{Posselt:2013xva}. More recently, observations in the FAINT mode---where pile-up effects are significantly reduced---have accumulated sufficient temporal coverage to confirm the rapid cooling of the Cas A NS at the $5\sigma$ level, assuming proper detector calibration~\cite{Posselt:2022pch}. The results from both observation modes are consistent and, when combined, suggest a cooling rate of $2.2\pm 0.3 \%$ over ten years, provided that the absorbing hydrogen column density is allowed to vary~\cite{Shternin:2022rti}. For a recent review on the Cas A NS cooling, see Ref.~\cite{Heinke:2025yxp}. Notably, this cooling rate can be naturally explained within the standard NS cooling scenario if the neutron triplet superfluid transition has recently occurred~\cite{Page:2010aw, Shternin:2010qi, Wijngaarden:2019tht, Shternin:2021fpt, Ho:2021hwy}. In this case, enhanced neutrino emission due to the breaking and reformation of neutron triplet Cooper pairs~\cite{Flowers:1976ux, Voskresensky:1986af, Voskresensky:1987hm} in the NS core would accelerate the cooling rate, making it consistent with the observed rate. See also Refs.~\cite{Yang:2011yg, Negreiros:2011ak, Blaschke:2011gc, Noda:2011ag, Sedrakian:2013pva, Blaschke:2013vma, Bonanno:2013oua, Leinson:2014cja, Taranto:2015ubs, Grigorian:2016leu, Tsiopelas:2020nzm, Hong:2020bxo, Leinson:2022qoc, Avila:2023rzj, Zhu:2024hex} for related discussions. 

In Ref.~\cite{Leinson:2014ioa}, it was proposed
that axion cooling could account for the observed cooling curve of the Cas A NS. Specifically, for the KSVZ axion, 
the analysis suggested
that $C_n^2/f_a^2 \simeq 1.6 \times 10^{-19}~\mathrm{GeV}^{-2}$ (\,$C_n$ denotes the axion-neutron coupling, as defined in Eqs.~\eqref{eq:axion_nucleon} and \eqref{eq:cn} below\,) is consistent with the observations. However, Ref.~\cite{Hamaguchi:2018oqw} pointed out that such a small value of $f_a$ would enhance axion emission from protons at earlier times---contribution not considered in Ref.~\cite{Leinson:2014ioa}. As a result, the predicted temperature at the current age of the Cas A NS would be lower than observed, leading to a constraint on the axion decay constant: $f_a \gtrsim 5 \times 10^8$~GeV for the KSVZ model and $f_a \gtrsim 7 \times 10^8$~GeV for the DFSZ model with $\tan \beta = 10$,\footnote{$\tan \beta$ is the ratio of the vacuum expectation values of the two Higgs doublets in the DFSZ axion model.} respectively. As a subsequent paper to Ref.~\cite{Leinson:2014ioa}, 
Ref.~\cite{Leinson:2021ety} revisited this analysis,
shifting the focus from explaining the cooling curve via axion cooling in Ref.~\cite{Leinson:2014ioa} to using the Cas A NS observations to constrain axion models. While 
the limit on the DFSZ axion in Ref.~\cite{Leinson:2021ety}
was comparable to that in Ref.~\cite{Hamaguchi:2018oqw}, 
the revised bound for the KSVZ axion in~\cite{Leinson:2021ety}
differed significantly, yielding $f_a > 3 \times 10^7$~GeV.

The purpose of this paper is to identify the source of this discrepancy. We find that the formula for axion emissivity from singlet proton Cooper pairs used in Ref.~\cite{Leinson:2021ety} differs in its dependence on the effective mass from those in other works~\cite{Keller:2012yr, Sedrakian:2015krq, Hamaguchi:2018oqw, Buschmann:2021juv}, leading to a significant difference in the total axion emission rate for the KSVZ axion. In Sec.~\ref{sec:calc}, we review the calculation of emissivity for this process, followed by an analysis in Sec.~\ref{sec:comparison} of how the difference in effective mass dependence affects the results. In Sec.~\ref{sec:eos}, we examine the uncertainty in axion emissivity due to the NS equation of state (EOS), which influences the calculation through the effective mass, Fermi momentum, and pairing energy gap. We also discuss the significance of this uncertainty on the discrepancy. Finally, Sec.~\ref{sec:conclusion} presents our conclusions.

\section{Axion emission from singlet Cooper pairs}
\label{sec:calc}

We begin by calculating the emissivity of axions from singlet proton Cooper pairs, a process known as pair-breaking and formation (PBF). In Ref.~\cite{Keller:2012yr}, this emissivity was derived using a transport equation and by evaluating the correlation functions of nucleon axial currents. Here, we revisit the calculation using the method of the Bogolyubov--Valatin transformations~\cite{Bogolyubov:1958km, Valatin:1958ja, Landau9}.  

The axion-nucleon interactions have the form 
\begin{equation}
  {\cal L}_{aNN}
=
  \sum_{N =p,n} \frac{C_N}{2 f_a}
  \bar{N} \gamma^\mu \gamma_5 N \, \partial_\mu a ~,
\label{eq:axion_nucleon}
\end{equation}
where $C_N$ ($N = p,n$) are the axion-nucleon couplings, whose size depends on axion models. For the KSVZ axion, the couplings are~\cite{GrillidiCortona:2015jxo} 
\begin{equation}
 C_p = -0.47(3)~, \qquad C_n =-0.02(3) ~. 
 \label{eq:cn}
\end{equation}
Notice that the axion-neutron coupling $C_n$ is highly suppressed in this model. 

To avoid unnecessary complications with subscripts, we perform our calculations exclusively for proton pairs. The results for neutrons can be easily obtained by replacing the corresponding quantities with those for neutrons.

In the presence of singlet Cooper pairs, the annihilation operators of quasi-particles on the paired ground state with momentum \(\bm{p}\) and spin \(s\), \(\hat{\alpha}_{\bm{p}, s}\), are related to those in the non-interacting Hamiltonian, \(\hat{a}_{\bm{p}, s}\), as~\cite{Bogolyubov:1958km, Valatin:1958ja, Landau9} 
\begin{align}
    \hat{\alpha}_{\bm{p}, +} &= u_p \hat{a}_{\bm{p}, +} - v_p \hat{a}^\dagger_{- \bm{p}, -} ~, \nonumber \\ 
    \hat{\alpha}_{\bm{p}, -} &= u_p \hat{a}_{\bm{p}, -} + v_p \hat{a}^\dagger_{- \bm{p}, +} ~,
    \label{eq:cantor}
\end{align}
with 
\begin{equation}
    u_p = \frac{1}{\sqrt{2}} 
    \biggl(1 + \frac{\eta_p}{\sqrt{\eta_p^2 + \Delta^2}}\biggr)^{\frac{1}{2}} ~, \qquad 
    v_p = \frac{1}{\sqrt{2}} \biggl(1 - \frac{\eta_p}{\sqrt{\eta_p^2 + \Delta^2}}\biggr)^{\frac{1}{2}} ~,
\end{equation}
where \(\Delta\) is the energy gap, and near the Fermi surface \(\eta_p\) has the form 
\begin{equation}
    \eta_p  \simeq  v_F (p- p_F) ~,
\end{equation}
where \(p_F\) and \(v_F\) are the Fermi momentum and velocity, respectively, and $p \equiv |\bm{p}|$.
%
%
The energy of a quasi-particle is given by 
\begin{equation}
    \epsilon_p = \sqrt{\eta_p^2 + \Delta^2} ~.
    \label{eq:epsilon_p}
\end{equation} 

The matrix elements of the proton field operator between the ground state \(|\Omega \rangle\) and the one-quasi-particle states \(| \bm{p}, s\rangle\) are\footnote{We use the Dirac representation for the four-component spinor coefficient functions. } 
\begin{align}
  \langle \Omega | p (x) | \bm{p}, s\rangle &= U_{s s'}(p) 
  \begin{pmatrix}
    \sqrt{p_0 + m^*}\,  \xi_{s'} \\ \sqrt{p_0 - m^*} \,\hat{\bm{p}} \cdot \bm{\sigma} \xi_{s'}
  \end{pmatrix}
  e^{-i p\cdot x} ~, \\[3pt] 
  \langle \Omega | \bar{p} (x) | \bm{p}, s\rangle &= 
  \left( \sqrt{p_0 + m^*}\,  \xi_{s'}^\dagger , \sqrt{p_0 - m^*}\,  \xi_{s'}^\dagger \,\hat{\bm{p}} \cdot \bm{\sigma} \right)
  V_{s^\prime s} (p) \, 
  e^{i p\cdot x} ~, 
\end{align}
where
%
%
%
\(\xi_s\) denotes a two-component spinor, \(\hat{\bm{p}} \equiv \bm{p}/|\bm{p}|\), \(\sigma^i\) (\(i = 1,2,3\)) are the Pauli matrices, 
the effective mass 
$m^*=p_{F}/v_{F}$, 
$p^0 \simeq m^* + \bm{p}^2/(2m^*)$,
and 
\begin{equation}
  U_{ss'} (p) = 
  \begin{pmatrix}
    u_p & 0 \\ 0 & u_p
  \end{pmatrix}
  ~, \qquad 
  V_{ss'} (p) = 
  \begin{pmatrix}
    0 & v_p \\ - v_p & 0 
  \end{pmatrix}
  ~. 
\end{equation} 

As it turns out, the emissivity of axions in the singlet PBF process vanishes in the non-relativistic limit, \(v_F \to 0\)~\cite{Keller:2012yr}.\footnote{This is the same as in the case of the neutrino PBF emission from singlet Cooper pairs via the nucleon axial current~\cite{Kaminker:1999ez}.} We therefore keep \(\mathcal{O} (v_F)\) terms in the amplitude calculation. The emissivity is given by 
\begin{align}
  \epsilon^S_a &= \frac{1}{2} \int \frac{d^3 \bm{p}}{(2\pi)^3 2p_0} \frac{d^3 \bm{p'}}{(2\pi)^3 2p_0'} \frac{d^3 \bm{q}}{(2\pi)^3 2q_0} \, q_0 f(p) f(p') \nonumber \\ 
  &\times  (2\pi)^4 \delta^3 (\bm{p} + \bm{p}' - \bm{q}) \delta (\epsilon_p + \epsilon_{p'} - q_0) \sum_{s, s'} \left| \mathcal{M} \right|^2 ~,
  \label{eq:emissivity} 
\end{align}
where \(q\), \(p\), and \(p'\) are the four-momenta of axion and quasi-particles, respectively, \(f(p)\) denotes the Fermi-Dirac distribution function, and 
\begin{equation}
  i \mathcal{M} =  - \frac{C_p}{2 f_a} q_\mu \langle \Omega | \bar{p} \gamma^\mu \gamma_5 p | \bm{p}, s ; \bm{p}' , s' \rangle ~. 
\end{equation}
The overall factor \(1/2\) in Eq.~\eqref{eq:emissivity} is to avoid double count in the momentum integral of quasi-particles. At \(\mathcal{O} (v_F^2)\), the square of the matrix element is obtained as 
\begin{align}
  \sum_{s, s'} \left| \mathcal{M} \right|^2 = \frac{C_p^2}{ 2 f_a^2} (2m^*)^2 
  \biggl[ & q_0^2 \frac{(\bm{p} - \bm{p}')^2}{(2m^*)^2} \left( u_p v_{p'} + u_{p'} v_p \right)^2 
  + \bm{q}^2   \left( u_p v_{p'} - u_{p'} v_p \right)^2 
  \nonumber \\ 
  &- 2q_0\, \bm{q} \cdot  \frac{(\bm{p} - \bm{p}')}{2m^*} \left( 
    u_p^2 v_{p'}^2 - u_{p'}^2 v_p^2 
   \right)
  \biggr] ~.
  \label{eq:Msq_axial_current}
\end{align}
The first, second, and third terms in the square bracket originate from the temporal, spatial, and temporal-spatial mixed parts,\footnote{In earlier studies~\cite{Kaminker:1999ez} on singlet PBF neutrino emission, the temporal-spatial mixed part of the squared matrix element of the axial-current was ignored, and this result was adopted in {\tt NSCool}\,\cite{nscool} for numerical simulations. We have verified that when applied to neutrino emission via the weak interaction, our evaluation of the squared axial-current matrix element in Eq.~\eqref{eq:Msq_axial_current} is consistent with the result in Ref.~\cite{Kolomeitsev:2008mc}, after accounting for medium corrections that suppress the vector current contribution. 
}
respectively---all of these contribute at the same order in \(v_F\)~\cite{Kolomeitsev:2008mc}, as we see below. 

We can readily perform the integral~\eqref{eq:emissivity} with respect to the axion momentum \(\bm{q}\), eliminating the momentum delta function with the momentum conservation \(\bm{q} = \bm{p}+ \bm{p}'\). For quasi-particle momenta, we note that for PBF emission in NSs, only narrow regions of momentum near the Fermi surface contribute to the integral. Hence, we can take \(|\bm{p}| = |\bm{p}'| = p_F\) in any smooth functions in the integrand. In particular, we can set 
\begin{equation}
  d^3 \bm{p} , ~ d^3 \bm{p}' \to p_F^2\, dp\, d\Omega_p , ~ p_F^2\, dp'\, d\Omega_{p'}  ~,
\end{equation}
where \(p = |\bm{p}|\), \(p' = |\bm{p}'|\), and \(d\Omega_p, d\Omega_{p'} \) are solid-angle elements. One of the angular integrals converts to a factor of \(4\pi\) because of the rotational invariance of the system. For the other one, the axial symmetry yields a factor of \(2 \pi\), and the remaining integral removes the delta function for energy conservation. Notice that since the axion energy in the PBF process is comparable to the size of the pairing gap, \(\Delta \ll p_F\) implies \(|\bm{q}| \ll |\bm{p}|, |\bm{p}'|\) , and thus \(\bm{p}' \simeq - \bm{p}\); only this anti-parallel region contributes to the angular integral in Eq.~\eqref{eq:emissivity}. From the kinematic condition, it follows that the discrepancy in the size of \(\bm{p}\) and \(\bm{p}'\) is restricted in the range \(|p-p'| \leq \epsilon_p + \epsilon_{p^\prime}\).

We are now left with the integral with respect to $p$ and $p'$. To simplify the expression, let us introduce the following dimensionless variables: 
\begin{align}
  x^{(\prime)} \equiv \frac{v_F (p^{(\prime)} - p_F)}{T} ~, \qquad 
  y \equiv \frac{\Delta }{T} ~, \qquad 
  z^{(\prime)} \equiv \frac{\epsilon_{p^{(\prime)}}}{T} ~,
\end{align}
with \(T\) the local temperature. We change the integral variables from \(p, p'\) to \(x, x'\), with the integral range restricted to 
\begin{equation}
  |x - x'| \leq v_F (z + z') ~.
\end{equation}
We can therefore regard $x - x'$ as an $\mathcal{O}(v_F)$ quantity. The terms in the square bracket in %
Eq.~\eqref{eq:Msq_axial_current}
are now computed as 
\begin{align}
  q_0^2 \frac{(\bm{p} - \bm{p}')^2}{(2m^*)^2} \left( u_p v_{p'} + u_{p'} v_p \right)^2  &= 4 v_F^2 T^2 y^2 ~, \\ 
  \bm{q}^2   \left( u_p v_{p'} - u_{p'} v_p \right)^2 &= T^2 \, \frac{y^2}{z^2} (x-x')^2 ~,  \\ 
  - 2q_0\, \bm{q} \cdot  \frac{(\bm{p} - \bm{p}')}{2m^*} \left( 
    u_p^2 v_{p'}^2 - u_{p'}^2 v_p^2  
   \right)&= - 2 T^2 \, \frac{y^2}{z^2} (x-x')^2 ~,
\end{align}
at the leading order in the \(v_F\) expansion. The integral~\eqref{eq:emissivity} is then expressed as
\begin{align}
  \epsilon_a^S = \frac{C_p^2}{16 \pi^3 f_a^2} m^{*2} T^5 y^2 \int_0^{\infty} dx \frac{z}{(e^z + 1)^2} \int_{-2v_Fz}^{2v_F z} d\xi 
  \biggl[4 v_F^2 + \frac{\xi^2}{z^2} - 2 \frac{\xi^2}{z^2}\biggr] ~,
\end{align} 
where we set \(\xi = x - x'\) and extend the upper limit of the \(x\) integral to infinity since this integral is rapidly convergent for large \(x\). After performing the \(\xi\) integral, we finally obtain 
\begin{align}
  \epsilon_a^S = \frac{2}{3} \cdot \frac{C_p^2}{\pi f_a^2} \cdot \biggl(\frac{m^* p_F}{\pi^2}\biggr) \,v_F^2\, T^5 \, F_s \biggl(\frac{\Delta}{T}\biggr)  ~,
  \label{eq:eas}
\end{align}
where 
\begin{equation}
  F_s (y) \equiv y^2 \int_0^{\infty} dx \frac{z^2}{(e^z + 1)^2} ~.
  \label{eq:fs}
\end{equation}

\section{Comparison of previous calculations}
\label{sec:comparison}

By noting 
\begin{equation}
  I_{aN}^S = F_s \biggl(\frac{\Delta^S_N (T)}{T}\biggr) ~,
\end{equation}
we find that the result obtained in the previous section is consistent with those in Refs.~\cite{Keller:2012yr, Sedrakian:2015krq}, which was used in the analysis in Ref.~\cite{Hamaguchi:2018oqw}. A useful fit function for the function~\eqref{eq:fs} is given in Ref.~\cite{Sedrakian:2015krq}: 
\begin{equation}
  F_s (z) = (a z^2 + cz^4) \sqrt{1 + fz} e^{-\sqrt{4z^2 + h^2} + h}~,
  \label{eq:ians}
\end{equation}
where $a = 0.158151$, $c = 0.543166$, $h = 0.0535359$, and $f = \pi/4 c^2 = 2.6621$. This fit was used in the numerical computation in Ref.~\cite{Hamaguchi:2018oqw}. Numerically, the emissivity~\eqref{eq:eas} is given by 
\begin{align}
  \epsilon^S_a \simeq 1.14 \times 10^{21} &\times \biggl(\frac{C_N}{2}\biggr)^2 \biggl(\frac{10^{10}~\mathrm{GeV}}{f_a}\biggr)^2 \biggl(\frac{m_N^*}{m_N}\biggr)^2 \biggl(\frac{v_{F,N}}{c}\biggr)^3 \biggl(\frac{T}{10^9~\mathrm{K}}\biggr)^5 \nonumber \\ 
  &\times F_s \biggl(\frac{\Delta^S_N (T)}{T}\biggr)~\mathrm{erg} \cdot \mathrm{cm}^{-3} \cdot \mathrm{s}^{-1} ~,
  \label{eq:semi}
\end{align}
which is again consistent with the expression given in Ref.~\cite{Keller:2012yr}. 

As seen in the case of the neutrino PBF emission via the axial current~\cite{Kolomeitsev:2008mc}, we expect a certain amount of medium corrections to the axion-nucleon interactions through the vertex and wave-function renormalization. In Ref.~\cite{Buschmann:2021juv}, this correction is estimated by a correction factor
\begin{equation}
  \gamma = \biggl[1 + \frac{1}{3} \biggl(\frac{m_N^*}{m_N}\biggr) \biggl(\frac{p_{F,n}}{1.68~\mathrm{fm}^{-1}}\biggr)\biggr]^{-1} ~. 
  \label{eq:corrfac}
\end{equation}
This can lead to an \(\mathcal{O} (10)\)\% reduction in the emissivity. Besides this factor and an overall numerical factor of $\simeq 1.5$, the semi-analytical expression shown in Ref.~\cite{Buschmann:2021juv} is consistent with Eq.~\eqref{eq:semi}. 

On the other hand, Ref.~\cite{Leinson:2021ety} shows a semi-analytic formula for the singlet axion PBF emissivity for proton\footnote{This is taken from the journal version and arXiv v3 of Ref.~\cite{Leinson:2021ety}. Different expressions are presented in other arXiv versions; in v1,
\begin{equation}
  Q_{pa}^{\mathrm{PBF}} = 1.55 \times 10^{40} g_{app}^2 \frac{p_{F,p}}{m^*_p} \biggl(\frac{m_p^*}{m_p}\biggr)^2 \biggl(\frac{T}{10^9~\mathrm{K}}\biggr)^5 \biggl(\frac{p_{F,p}}{m_pc}\biggr)^2 \biggl[\biggl(\frac{m_p^*}{m_p}\biggr)^2 + \frac{11}{42}\biggr] F_2 \biggl(\frac{T}{T_{cp}}\biggr) ~\frac{\mathrm{erg}}{\mathrm{cm}^3 \cdot \mathrm{s}} ~,
  \label{eq:qpav1}
\end{equation}
and in v2 
\begin{equation}
  Q_{pa}^{\mathrm{PBF}} = 1.55 \times 10^{40} g_{app}^2 \frac{p_{F,p}}{m^*_p} \biggl(\frac{m_p^*}{m_p}\biggr)^2 \biggl(\frac{T}{10^9~\mathrm{K}}\biggr)^5 \biggl(\frac{p_{F,p}}{m_pc}\biggr)^2  \frac{6}{7} F_2 \biggl(\frac{T}{T_{cp}}\biggr) ~\frac{\mathrm{erg}}{\mathrm{cm}^3 \cdot \mathrm{s}} ~.
  \label{eq:qpav2}
\end{equation}
}  
\begin{equation}
    Q_{pa}^{\mathrm{PBF}} = 1.55 \times 10^{40} g_{app}^2 \biggl(\frac{m_p^*}{m_p}\biggr)^2 \biggl(\frac{T}{10^9~\mathrm{K}}\biggr)^5 \biggl(\frac{p_{F,p}}{m_pc}\biggr)^3 \frac{6}{7} F_2 \biggl(\frac{T}{T_{cp}}\biggr) ~\frac{\mathrm{erg}}{\mathrm{cm}^3 \cdot \mathrm{s}} ~,
    \label{eq:qpa}
\end{equation}
where $g_{app} = C_p m_p/f_a$ and 
\begin{align}
    F_2 (T/T_c) &= \frac{\Delta_N^2}{T^2} \int_0^{\infty} dx \, \frac{z^2}{(\exp z + 1)^2} ~,
\end{align}
with $z = \sqrt{x^2 + \Delta_N^2/T^2}$; \textit{i.e.,} \(F_2 = F_s\). We observe that, apart from an overall factor, this expression exhibits a different dependence on the effective mass compared to Eq.~\eqref{eq:eas}; the emissivity in Eq.~\eqref{eq:qpa} is proportional to \((m_p^*)^2 p_{F,p}^3\),\footnote{The emissivities in arXiv v1 and v2 of \cite{Leinson:2021ety} (Eq.~\eqref{eq:qpav1} and Eq.~\eqref{eq:qpav2}, respectively) also exhibit different dependencies on the effective mass. } whereas Eq.~\eqref{eq:eas} is proportional to \(p_{F,p}^3/m_p^*\). We are unable to determine the origin of this discrepancy since the derivation of the axion emissivity was not provided in Ref.~\cite{Leinson:2021ety}.  Given that the effective mass in the NS core can be significantly smaller than that in the vacuum, this difference could lead to a substantial deviation in the emissivity.

\begin{figure}
  \centering
  \includegraphics[height=85mm]{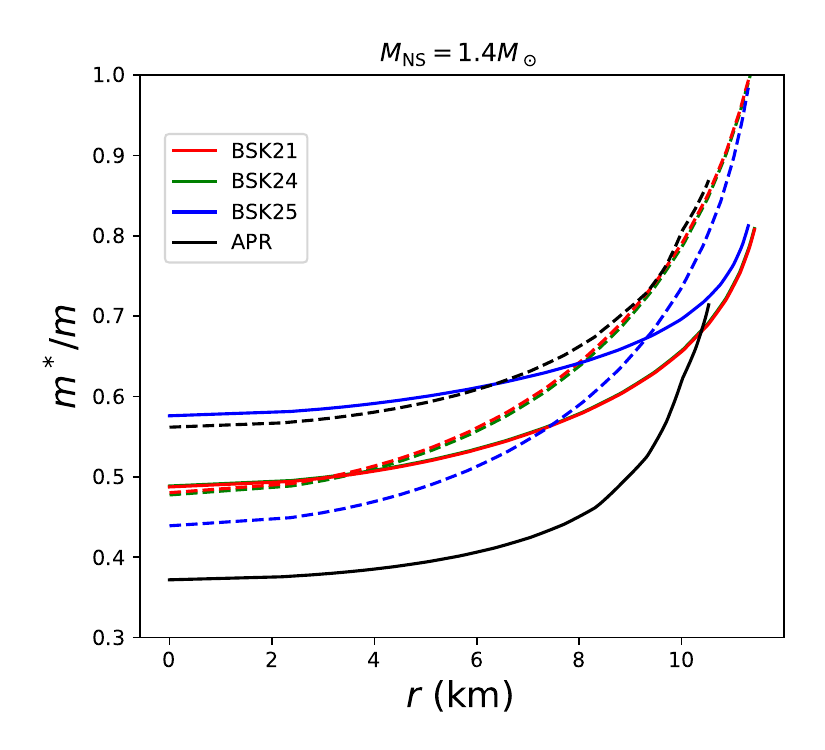}
  \caption{
    The ratio \(m^*_N/m_N\) as functions of the distance from the center of the NS, \(r\). The solid and dashed lines are for proton and neutron, respectively. The black, red, green, and blue lines correspond to NSs with a mass of \(1.4M_\odot\), constructed using the APR EOS~\cite{Akmal:1998cf}, the BSK21 EOS~\cite{Goriely:2010bm}, the BSK24 EOS~\cite{Pearson:2018tkr}, and the BSK25 EOS~\cite{Pearson:2018tkr}, respectively. 
  }  
  \label{fig:mstm_vs_r}
\end{figure}

To illustrate this, in Fig.~\ref{fig:mstm_vs_r}, we show the ratio \(m^*_N/m_N\) as a function of the distance \(r\) from the center of the NS, where the solid and dashed lines are for proton and neutron, respectively. The black and red lines correspond to NSs with a mass of \(1.4M_\odot\), constructed using the APR EOS~\cite{Akmal:1998cf} (adopted in Ref.~\cite{Hamaguchi:2018oqw}) and the BSK21 EOS~\cite{Goriely:2010bm,Pearson:2011zz} (adopted in Ref.~\cite{Leinson:2021ety}). 
We also present effective masses for more recent EOSs provided in Ref.~\cite{Pearson:2018tkr}.%
\footnote{For density above $10^{6}~{\rm g/cm^3}$ corresponding to the core and the inner crust of a NS, we obtain the BSK EOSs from the fitting programs provided in \cite{BSKsite}. For lower densities in the outer crust, the EOSs are read from \cite{Pearson:2011zz,Pearson:2018tkr}. To build the NS profile, we modified and ran the Tolman-Oppenheimer-Volkov equation solver and the main program in the {\tt NSCool}\,\cite{nscool} package.}
The results for the BSK24 and BSK25 EOSs are shown in green and blue, respectively. Notably, the BSK24 result (green) is almost identical to that of BSK21 (red), and this similarity persists in the following analysis. We do not include results for BSK22 and BSK26~\cite{Pearson:2018tkr}. The BSK22 EOS allows the direct URCA process to operate in a NS with a mass \(\gtrsim 1.151M_\odot\)\,\cite{Pearson:2018tkr}, which would render the temperature of Cas A---estimated to have a mass of \(1.55 \pm 0.25M_\odot\)~\cite{Shternin:2022rti}---too low at the time of observation. On the other hand, the BSK26 EOS is inconsistent with the mass-radius measurements of PSR~J0030 and PSR~J0740 based on {\it NICER} data~\cite{Buschmann:2021juv,Miller:2021qha}.

As can be seen in Fig.~\ref{fig:mstm_vs_r},
in the core region \(m_p^*/m_p\) can be as low as \(\simeq 0.5\). Consequently, the factor of \((m_p^*)^3\) leads to almost an order-of-magnitude difference in emissivity between Ref.~\cite{Hamaguchi:2018oqw} and Ref.~\cite{Leinson:2021ety}. 

\begin{figure}
  \centering
  \subcaptionbox{\label{fig:BSK21pbf_Q} Axion emissivity }{
  \includegraphics[width=0.48\columnwidth]{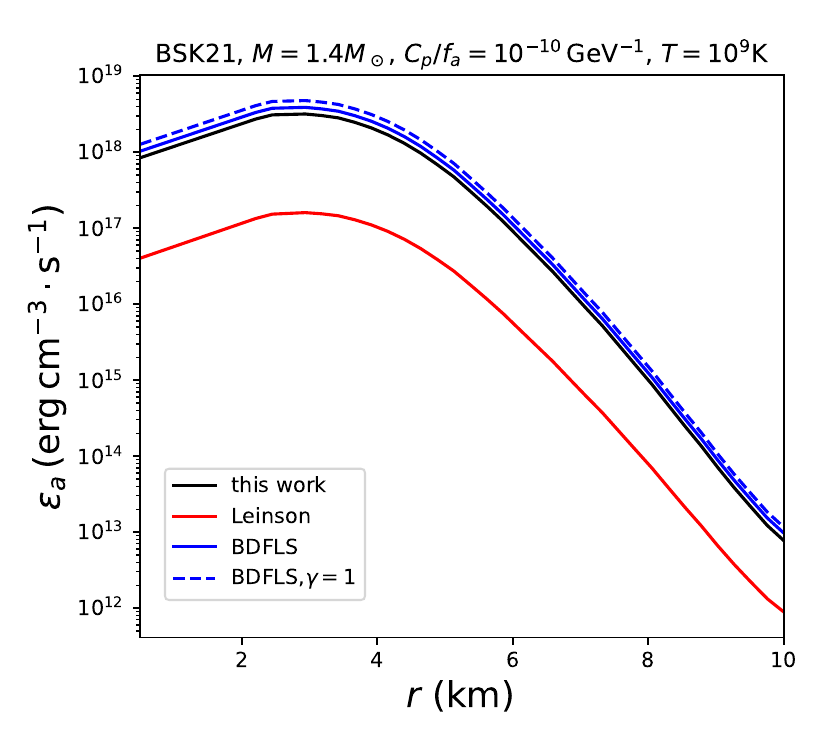}}
  \subcaptionbox{\label{fig:BSK21pbf_AQ} Emission rate in a spherical shell}{
  \includegraphics[width=0.48\columnwidth]{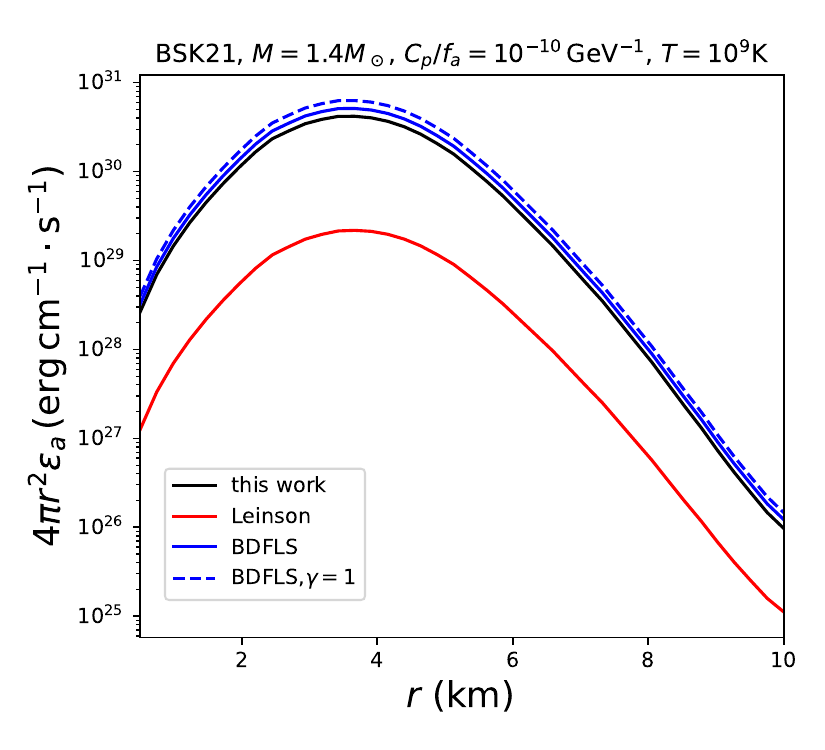}}
\caption{a) The axion emissivity and b) the emission rate in a spherical shell from proton singlet pairings as functions of the distance from the center of a \(1.4M_\odot\) NS constructed with the BSK21 EOS and the CCDK proton pairing gap model~\cite{Chen:1993bam}, where we set \(T = 10^9\)~K and \(C_p/f_a = 10^{-10}~\mathrm{GeV}^{-1}\). The black solid, red solid, blue solid, and blue dashed lines correspond to Eq.~\eqref{eq:eas}, Eq.~\eqref{eq:qpa}, the expression given in Ref.~\cite{Buschmann:2021juv}, and that in Ref.~\cite{Buschmann:2021juv} with \(\gamma = 1\), respectively. 
} 
  \label{fig:BSK21pbf}
\end{figure}

We compare our results with the previous calculations in Fig.~\ref{fig:BSK21pbf}. Figures~\ref{fig:BSK21pbf_Q} and \ref{fig:BSK21pbf_AQ} show the axion emissivity, \(\epsilon^S_a\), and the emission rate in a spherical shell, \(4\pi r^2 \epsilon^S_a\), respectively, as functions of the distance from the center of a \(1.4M_\odot\) NS constructed with the BSK21 EOS and the CCDK proton pairing gap model~\cite{Chen:1993bam}. We set \(T = 10^9\)~K and \(C_p/f_a = 10^{-10}~\mathrm{GeV}^{-1}\). The black solid line represents our result, as described in Sec.~\ref{sec:calc}, which is consistent with those used in Refs.~\cite{Keller:2012yr, Sedrakian:2015krq,Hamaguchi:2018oqw}. The blue solid (dashed) line corresponds to the semi-analytical expression in Ref.~\cite{Buschmann:2021juv} with (without) the correction factor \(\gamma\) in Eq.~\eqref{eq:corrfac}. The red solid line shows the semi-analytical result from the journal version and arXiv v3 of Ref.~\cite{Leinson:2021ety}. It turns out that our result, given in Eq.~\eqref{eq:eas}, agrees well with that of Ref.~\cite{Buschmann:2021juv}. In particular, the effect of the medium correction is small, indicating that the theoretical uncertainty in our calculation is well controlled. In contrast, the emissivity in Ref.~\cite{Leinson:2021ety} is found to be significantly lower than the others. This discrepancy is attributed to the difference in the dependence on the effective mass. 

As discussed in Ref.~\cite{Hamaguchi:2018oqw}, for the KSVZ axion, the proton PBF is the dominant axion emission process because of the suppressed axion-neutron coupling (see Eq.~\eqref{eq:cn}). As a result, a lower emissivity of the proton PBF leads to a weaker limit on the axion decay constant. Indeed, the axion emission luminosity of the proton PBF shown in Ref.~\cite{Leinson:2021ety} is significantly lower than that in Ref.~\cite{Hamaguchi:2018oqw}. This indicates that the weaker constraint on the KSVZ axion reported in Ref.~\cite{Leinson:2021ety} stems from an underestimated proton PBF emissivity, which in turn arises from an incorrect effective mass dependence in the axion emissivity formula used in Ref.~\cite{Leinson:2021ety}.

\section{EOS dependence}
\label{sec:eos}

\begin{figure}
  \centering
  \subcaptionbox{\label{fig:kfp_vs_r} Proton Fermi momentum}{
  \includegraphics[width=0.48\columnwidth]{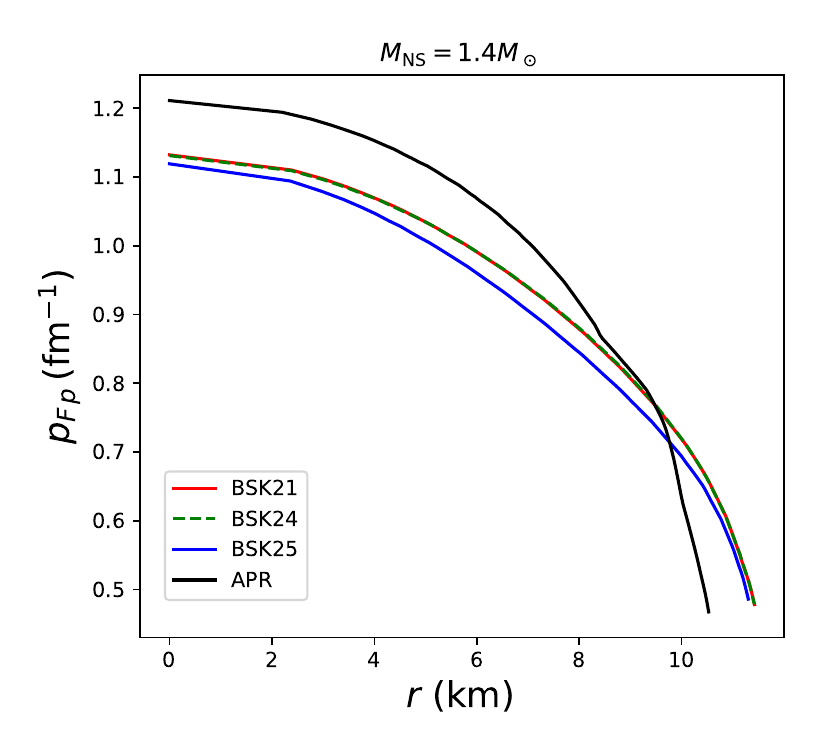}}
  \subcaptionbox{\label{fig:Deltap_vs_r} proton singlet pairing gap}{
  \includegraphics[width=0.48\columnwidth]{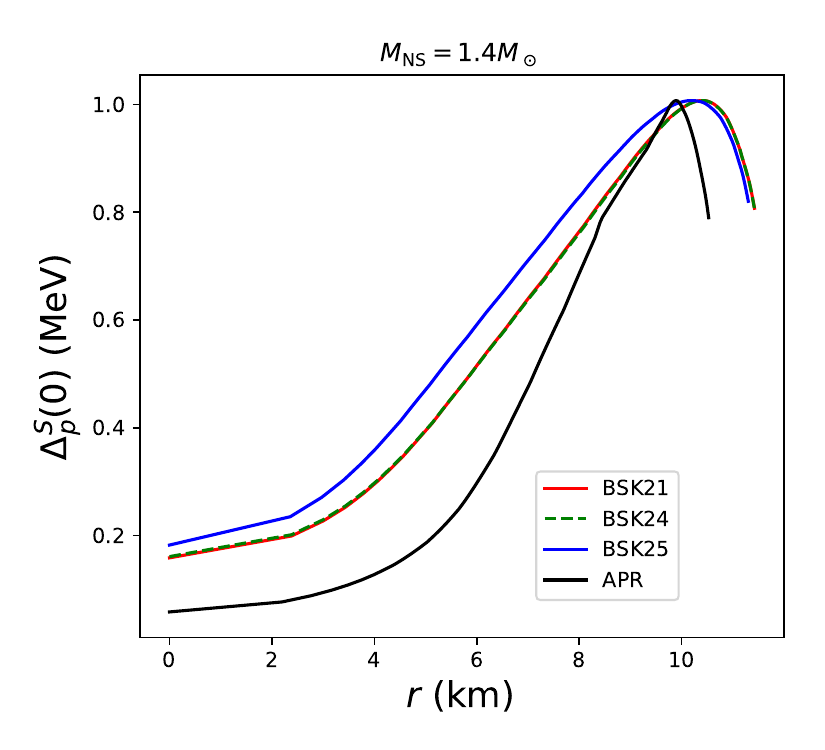}}
\caption{a) The proton Fermi momentum as a function of the distance from the NS center; b) The proton singlet pairing gap profile at $T=0$ obtained with the CCDK gap model.
\label{fig:kfpandDelp}} 
\end{figure}

As evident from Eq.~\eqref{eq:eas}, the emissivity of axions depends not only on the effective mass but also on the Fermi momentum and the pairing gap, both of which are influenced by the choice of NS EOS, as shown in Fig.~\ref{fig:kfpandDelp}. In Fig.~\ref{fig:kfp_vs_r}, we show the proton Fermi momenta $p_{F,p}$ as a function of the distance from the NS center, computed for different EOSs. For $r\lesssim 9$~km, the proton Fermi momenta---and consequently, the proton densities---exhibit only minor variations among the BSK EOSs, whereas those for the APR EOS are \(\sim 10\%\) higher than the others. On the other hand, in the lower density region at $r\gtrsim 9$~km, the proton Fermi momentum for the APR EOS is much lower than that for the BSK21, 24, and 25 EOSs. It is, however, worth noting that axion emission in this region is highly suppressed compared to the NS core, as can be seen in Fig.~\ref{fig:EOSs_Q_vs_r} below.

Fig.~\ref{fig:Deltap_vs_r} shows the profile of the proton singlet pairing gap $\Delta_p^S$ at $T=0$, obtained with the CCDK gap model for various EOSs. The variations in $\Delta_p^S$ across different EOSs originate from the dependence of the pairing gap on $p_{F,p}$. In the core region, the pairing gap for the APR EOS differs by an \(\mathcal{O}(1)\) factor from those for BSK EOSs, which significantly alters the emissivity due to its exponential dependence on $\Delta_p^S$, as seen in Eq.~\eqref{eq:fs}. 

\begin{figure}
  \centering
  \subcaptionbox{\label{fig:EOSspbf_Q} Axion emissivity}{
  \includegraphics[width=0.48\columnwidth]{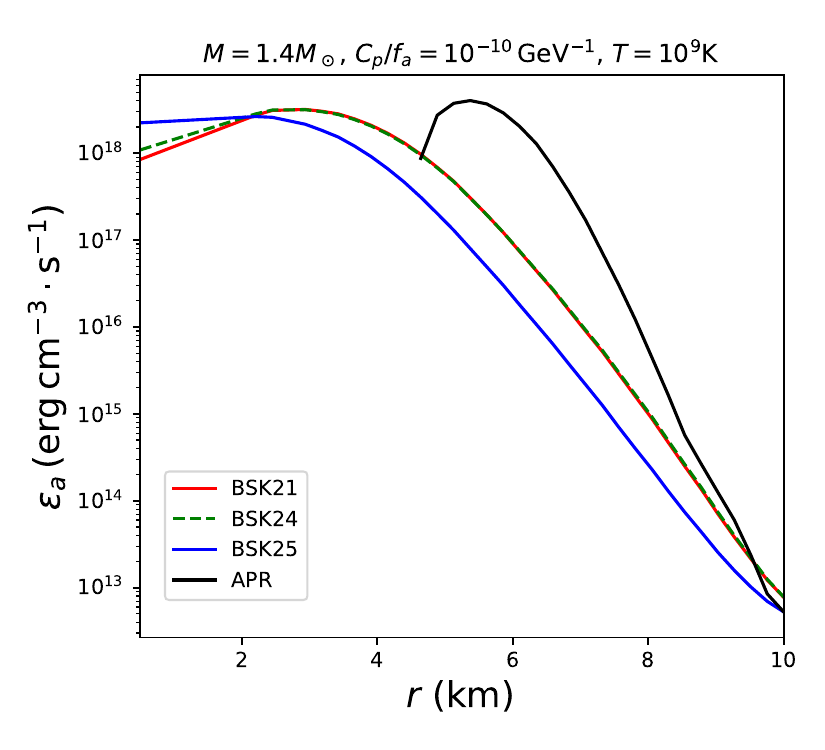}}
  \subcaptionbox{\label{fig:EOSspbf_AQ} Axion emission within a spherical shell}{
  \includegraphics[width=0.48\columnwidth]{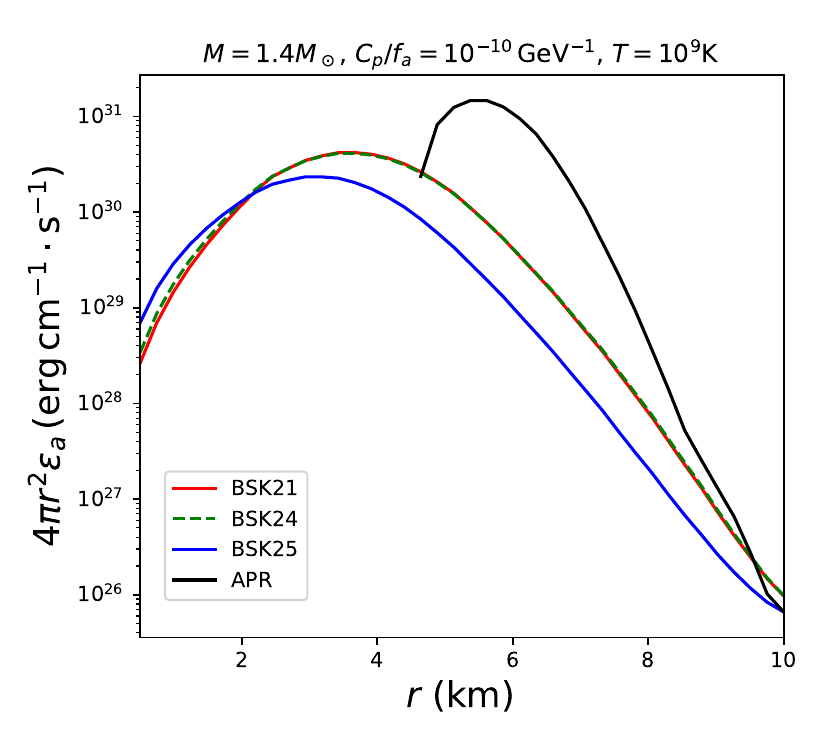}}
\caption{a) The axion emissivity and b) the axion emission rate in a spherical shell from proton singlet pairings as functions of the distance from the center of a \(1.4M_\odot\) NS constructed with various EOSs, where we use the CCDK proton pairing gap model~\cite{Chen:1993bam} and set \(T = 10^9\)~K and \(C_p/f_a = 10^{-10}~\mathrm{GeV}^{-1}\).
} 
  \label{fig:EOSs_Q_vs_r}
\end{figure}

This effect is illustrated in Fig.~\ref{fig:EOSs_Q_vs_r},
where we plot the axion emissivity in Fig.~\ref{fig:EOSspbf_Q} and the axion emission rate within a spherical shell in Fig.~\ref{fig:EOSspbf_AQ}, both from proton singlet pairings, as functions of the distance from the center of a \(1.4M_\odot\) NS constructed with various EOSs. We use the CCDK proton pairing gap model~\cite{Chen:1993bam} and set \(T = 10^9\)~K and \(C_p/f_a = 10^{-10}~\mathrm{GeV}^{-1}\). For the APR EOS, plotted with the black line, protons are unpaired at $r\lesssim 4$~km because of the small proton pairing gap in this region, as shown in Fig.~\ref{fig:Deltap_vs_r}. As a result, the proton PBF process is inactive in this region for the APR EOS. 

\begin{figure}
  \centering
  \subcaptionbox{\label{fig:EOSspbf_Lratio}}{
  \includegraphics[width=0.48\columnwidth]{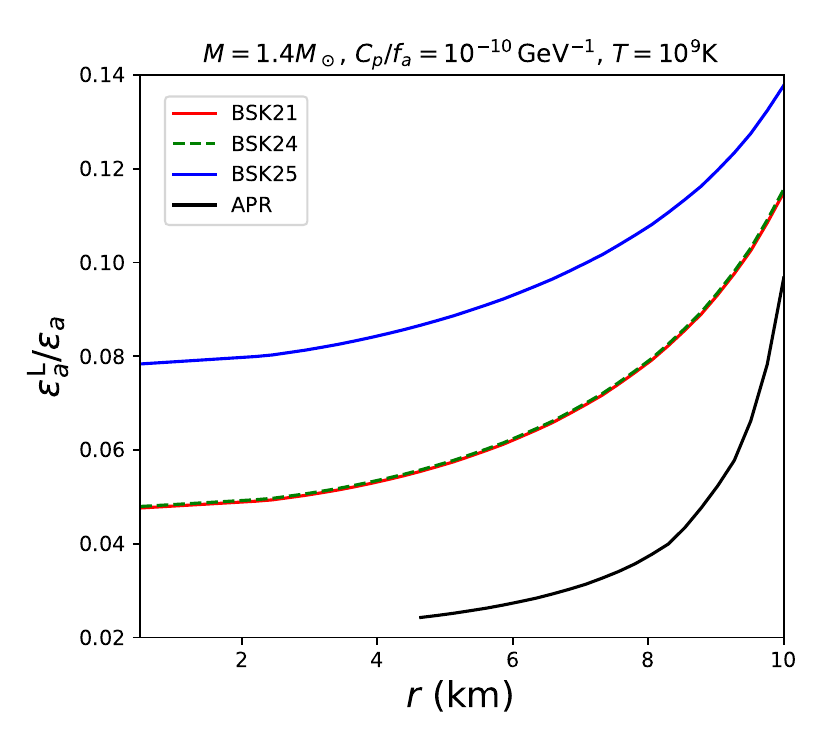}
  }
  \subcaptionbox{\label{fig:EOSspbf_Bratio} }{
  \includegraphics[width=0.48\columnwidth]{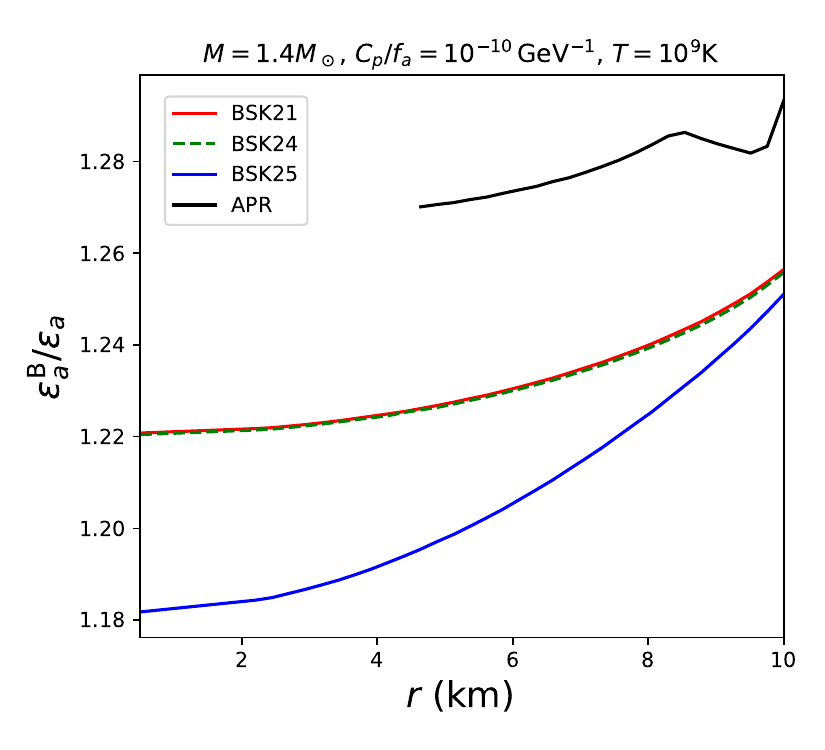}}
\caption{The ratio of emissivities calculated using a) Eq.~\eqref{eq:qpa} from Ref.~\cite{Leinson:2021ety} and b) expression in Ref.~\cite{Buschmann:2021juv} with the medium correction factor, to our result in Eq.~\eqref{eq:semi}, as functions of the distance from the center of a \(1.4M_\odot\) NS constructed with various EOSs. Other parameters are the same as those in Fig.~\ref{fig:EOSs_Q_vs_r}.
} 
  \label{fig:EOSs_Qratio_vs_r}
\end{figure}

In Fig.~\ref{fig:EOSs_Qratio_vs_r}, we present the ratios of proton-PBF axion emissivities obtained from expressions in the literature to the result re-derived in this work, plotted as functions of the distance from the center of a \(1.4M_\odot\) NS constructed with various EOSs. In this figure, $\epsilon_a$ is calculated using the semi-analytical expression in Eq.~\eqref{eq:semi}, while $\epsilon_a^{\rm L}$ in Fig.~\ref{fig:EOSspbf_Lratio} is obtained from Eq.~\eqref{eq:qpa} adopted from Ref.~\cite{Leinson:2021ety} and $\epsilon_a^{\rm B}$ in Fig.~\ref{fig:EOSspbf_Bratio} is calculated from Eq.~(S8) in Ref.~\cite{Buschmann:2021juv} including the medium correction factor $\gamma$. Figure~\ref{fig:EOSspbf_Lratio} shows that the ratio $\epsilon^{\rm L}_a/\epsilon_a$ is much smaller than unity and highly sensitive to the choice of the EOS due to the additional $(m_p^*/m_p)^3$ dependence in Eq.~\eqref{eq:qpa} compared to Eq.~\eqref{eq:semi}. As we have seen in Fig.~\ref{fig:mstm_vs_r}, the proton effective mass can be considerably lower than its vacuum mass and varies significantly among different EOSs, leading to a discrepancy between the emissivities obtained in Ref.~\cite{Leinson:2021ety} and those in Ref.~\cite{Hamaguchi:2018oqw} by more than an order of magnitude, with large uncertainties arising from the choice of EOSs. On the other hand, as shown in Fig.~\ref{fig:EOSspbf_Bratio}, the difference between $\epsilon^{\rm B}_a$ and $\epsilon_a$ is just an \(\mathcal{O}(10)\)\% level and exhibits much smaller variations across different EOSs. This is because the EOS dependence in the ratio $\epsilon^{\rm B}_a/\epsilon_a$ enters only through the medium correction factor $\gamma$ in $\epsilon^{\rm B}_a$, as described in Eq.~\eqref{eq:corrfac}, which is far less sensitive to the EOS than the $(m_p^*/m_p)^3$ factor in $\epsilon^{\rm L}_a/\epsilon_a$.

\section{Conclusion}
\label{sec:conclusion}

In this study, we have re-derived the emissivity of axions from singlet proton Cooper pairs in NSs and compared it with previous results in the literature. Our re-derived expression, which was used in Ref.~\cite{Hamaguchi:2018oqw} and consistent with those in Refs.~\cite{Keller:2012yr, Sedrakian:2015krq}, is in close agreement with the results in Ref.~\cite{Buschmann:2021juv} but differs significantly from those in Ref.~\cite{Leinson:2021ety}. This discrepancy, combined with different choices of the NS EOS, can explain the differences in the proton-PBF emissivities of the KSVZ axion for the Cas A NS in Refs.~\cite{Hamaguchi:2018oqw} and Ref.~\cite{Leinson:2021ety}. We have also investigated the impact of different EOS choices on axion emissivity and found that the large discrepancy persists regardless of the choice of EOSs.

As shown in Fig.~\ref{fig:mstm_vs_r}, the effective mass of not only protons but also neutrons depends on the choice of EOS. This, together with the EOS dependence of the Fermi momentum and pairing gaps of neutrons, also affects the total axion emissivity as well as that of neutrinos. All of these factors collectively  affect the constraints on axion-nucleon couplings derived from the Cas A NS. We will update these constraints by systematically incorporating EOS dependence and utilizing the latest data of the temperature observations of the Cas A NS in future work.

\section*{Acknowledgments}

We thank Andrew J. Long for useful correspondence.
This work was supported by JSPS KAKENHI Grant Numbers 24H02244 (KH), 24K07041 (KH), and 21K13916 (NN).


\bibliographystyle{utphysmod}
\bibliography{ref}


\end{document}